\documentclass[prd]{revtex4}

\usepackage{graphicx}
\usepackage{dcolumn}
\usepackage{bm}

\newcommand{\Rea}{\ensuremath{\mathrm{Re}} }
\newcommand{\Imm}{\ensuremath{\mathrm{Im}} }
\newcommand{\To}{\ensuremath{\rightarrow}}
\newcommand{\kz}{\ensuremath{K^0} }
\newcommand{\kzb}{\ensuremath{\overline{K}^0} }
\newcommand{\ks}{\ensuremath{K_S} }
\newcommand{\kl}{\ensuremath{K_L} }

\newcommand{\beps}{\ensuremath{\overline{\epsilon}} }
\newcommand{\epseps}{\ensuremath{\epsilon'/\epsilon}}

\begin{document}

\title{On the direct $CP$ violation parameter $\epsilon'$}

\author{M. S. Sozzi}
\email{marco.sozzi@sns.it}
\affiliation{Scuola Normale Superiore \\
Piazza dei Cavalieri 7 - 56126 Pisa, Italy}

\date{\today}

\begin{abstract}
We review different definitions of the $\epsilon'$ parameter describing direct 
CP violation in neutral kaon decays, which was precisely measured in recent 
experiments, and point out the inconsistency of some of them due to a CPT
constraint. 
The proper comparison of the experimental results to the theoretical 
computations is discussed.
\end{abstract}


\maketitle

\section{Introduction}

Among the important achievements of experimental physics in the past few 
years, the clarification of the long-standing puzzle concerning the existence 
of direct $CP$ violation in nature has an important place.
The definitive proof that $CP$ violation is indeed present in the decay 
amplitudes of the long-lived neutral kaon to $\pi\pi$ final states 
\cite{NA48_eprime} \cite{KTeV_eprime}, as expressed by the small but non-zero 
parameter $\epsilon'$, is the culmination of an experimental program which 
started 30-years ago, right after the discovery of $CP$ violation \cite{CPV}, 
and was strongly pursued since then, with several dedicated efforts in the 
past two decades (see \emph{e.g.} \cite{Cimento} for a recent review).

The deep meaning of such a result lies in the indication that $CP$ violation, 
being present also in its direct form as expected from the current CKM 
paradigm, is truly an ubiquitous feature of weak interactions, not limited to 
the peculiar $\kz-\kzb$ system as the super-weak \emph{ansatz} 
\cite{Superweak} would suggest. 
This fact was experimentally confirmed just a few years after the definitive 
proof of direct $CP$ violation, when $CP$ violation in the neutral $B$ meson 
system was measured with significant statistics at the $B$-factories 
\cite{Babar} \cite{Belle}.

While the main importance of the result is expressed by the fact that 
$\epsilon' \neq 0$ (with a significance which at present exceeds 7 standard 
deviations), regardless of its exact value, one should not oversee the fact 
that this parameter is now measured at the $\sim 15\%$ level, and improvements 
on the precision are expected when the final result from the full KTeV 
statistics and data from KLOE will be available.

Although the theoretical control of the $\epsilon'$ parameter is still poor 
at present, the situation is expected to improve in the future, particularly 
due to progress in lattice QCD computations, and the $CP$-violating parameters 
of the $K$ meson system could also acquire more value as \emph{quantitative} 
tests of the Standard Model, as well as constraints on models of New Physics.

In this perspective, it seems appropriate to establish a clean framework in 
which experimental measurements are to be compared with the theoretical 
predictions and among themselves, while avoiding possible confusion which 
could arise due to the existence of several alternative formulations of the 
phenomenological description.

The plan of the paper is as follows: in section \ref{sec:definitions} we 
briefly review a simple and consistent parameterization of $CP$-violation  in 
the $K$ system, and in section \ref{sec:others} we compare it with other 
formulations appearing in the literature, pointing out in section 
\ref{sec:consistency} some inconsistencies which are usually overlooked.
We then summarize in section \ref{sec:averages} the experimental knowledge 
on the $\epsilon'$ parameter. Finally, section \ref{sec:conclusions} presents 
our conclusions.

\section{$CP$-violating parameters in the neutral $K$ system}
\label{sec:definitions}

The phenomenological description of $CP$ violation in the neutral kaon system 
has its roots in the classic seminal papers by Wu, Yang and Lee \cite{WuYang} 
\cite{LeeWu}.
Such description involves the two complex parameters $\epsilon$ and 
$\epsilon'$, intended to parameterize respectively the so-called ``indirect'' 
$CP$ violation, defined \cite{Nir} as that occurring in the $|\Delta S| = 2$ 
virtual transitions described by the effective Hamiltonian in the $\kz-\kzb$ 
sub-space, and the ``direct'' $CP$ violation occurring in the physical 
$|\Delta S| = 1$ decay amplitudes to real final states such as $\pi^+ \pi^-$ 
or $\pi^0 \pi^0$.

In discussing $CP$ violation, care should be taken in considering which 
parameters are unphysical because their value depends on the arbitrary choice 
of the phase for the state vectors representing the different particles; 
indeed, there is a considerable amount of literature concerning the proper 
definition of rephasing-invariant parameters in the kaon system (see 
\emph{e.g.} \cite{Wu} \cite{Chou} \cite{Tsai}).

We now introduce some definitions \cite{Chau} \cite{Branco}; we will assume 
the validity of $CPT$ symmetry in the following, unless explicitly indicated 
otherwise. 

The $CP$-violating measurable ratios of amplitudes for decays of neutral 
kaons into a final $CP$ eigenstate $|f\rangle$ with eigenvalue $CP$=+1 are
\begin{equation}
  \eta_f \doteq
  \frac{\langle f | \mathcal{T} | \kl \rangle}
       {\langle f | \mathcal{T} | \ks \rangle}
  \frac{\langle \kz | \ks \rangle}{\langle \kz | \kl \rangle}
\end{equation}
where $\mathcal{T}$ is the transition matrix of weak interactions, and 
the second factor makes the $\eta_f$ parameter invariant under rephasing of 
both the $|\kz\rangle, |\kzb\rangle$ and $|\ks\rangle, |\kl\rangle$ state 
vectors \cite{Kayser}; such factor is often omitted, implicitly making the 
choice of a phase convention in which its value is 1.

In an analog way, rephasing-invariant amplitude ratios can be defined for 
other (non-observable) $CP$-even final states, such as those with two pions in 
a definite isospin eigenstate with eigenvalue $I$:
\begin{equation}
  \eta_I \doteq 
  \frac{\langle (\pi \pi)_I | \mathcal{T} | \kl \rangle}
       {\langle (\pi \pi)_I | \mathcal{T} | \ks \rangle}
  \frac{\langle \kz | \ks \rangle}{\langle \kz | \kl \rangle}
\end{equation}
and the usual $\epsilon$ parameter is defined as
\begin{equation}
  \epsilon \doteq \eta_0
\label{eq:eps}
\end{equation}

The quantity 
\begin{equation}
  \omega \doteq 
  \frac{\langle (\pi \pi)_{I=2} | \mathcal{T} | \ks \rangle}
       {\langle (\pi \pi)_{I=0} | \mathcal{T} | \ks \rangle}
\label{eq:omega}
\end{equation}
parameterizes the violation of the so-called (and poorly understood) 
$\Delta I = 1/2$ rule.
The modulus of such parameter can be extracted from the $\pi\pi$ decay rates 
of $K^+$ (which require $\Delta I > 1/2$) and \ks (see \emph{e.g.} 
\cite{Branco}), and is $|\omega| \simeq 0.045$.

The direct $CP$ violation parameter is finally
\begin{equation}
  \epsilon' \doteq \frac{1}{\sqrt{2}} \left[
  \frac{\langle (\pi \pi)_{I=2} | \mathcal{T} | \kl \rangle}
       {\langle (\pi \pi)_{I=0} | \mathcal{T} | \ks \rangle} - 
  \frac{\langle (\pi \pi)_{I=0} | \mathcal{T} | \kl \rangle}
       {\langle (\pi \pi)_{I=0} | \mathcal{T} | \ks \rangle} 
  \frac{\langle (\pi \pi)_{I=2} | \mathcal{T} | \ks \rangle}
       {\langle (\pi \pi)_{I=0} | \mathcal{T} | \ks \rangle} \right] 
  \frac{\langle \kz | \ks \rangle}{\langle \kz | \kl \rangle} = 
  \frac{\omega}{\sqrt{2}} (\eta_2 - \eta_0)
\label{eq:epsprime}
\end{equation}

All the above parameters are explicitly invariant under rephasing of both
$|\kz\rangle,|\kzb\rangle$ and $|\ks\rangle,|\kl\rangle$, and agree with those 
used in a significant number of papers describing the phenomenology of $CP$ 
violation in neutral kaon decays.

$CPT$ symmetry and the symmetry of strong interactions under time reversal 
(plus unitarity) allow to write the decay amplitudes for $\kz,\kzb$ into 
$\pi\pi$ states of definite isospin as 
\begin{eqnarray}
  A_I \doteq \langle (\pi \pi)_I | \mathcal{T} | \kz \rangle = 
  a_I e^{i\delta_I} &&
  \overline{A}_I \doteq \langle (\pi \pi)_I | \mathcal{T} | \kzb \rangle = 
  a_I^* e^{i\delta_I}
\end{eqnarray}
factorizing the $CP$-even $\pi\pi$ scattering phase $\delta_I$ by the 
Fermi-Watson theorem.

It can be easily seen \cite{Nir} that the parameter $\epsilon$ contains both 
indirect and direct $CP$ violation, and the separation of the two parts has no 
physical meaning, depending on the choice of phase convention:
\begin{equation}
  \epsilon = \frac{\beps + i\xi_0}{1+i\beps \, \xi_0}
\end{equation}
where $\beps$ is the (phase-convention dependent) mixing parameter which 
describes the $\kz,\kzb$ components in the physical eigenstates of the 
effective Hamiltonian contributing to indirect $CP$ violation:
\begin{eqnarray}
  & | \ks \rangle = \frac{1}{\sqrt{2(1+|\beps|^2)}}
    \left[ (1+\beps) |\kz \rangle + (1-\beps) |\kzb \rangle \right] \\
  & | \kl \rangle = \frac{1}{\sqrt{2(1+|\beps|^2)}}
    \left[ (1+\beps) |\kz \rangle - (1-\beps) |\kzb \rangle \right] 
\end{eqnarray}
(having arbitrarily fixed the relative phase between $|\ks\rangle$ and
$|\kl\rangle$), and $\xi_I \doteq \Imm(a_I)/\Rea(a_I)$ is a measure of the 
(unphysical) weak phase of the decay amplitude of \kz into a $\pi\pi$ state 
of isospin $I$.
It should be reminded that the unphysical parameter $\beps$ can be very large 
even if $CP$ violation itself is a small effect, \emph{i.e.} one can choose a 
phase convention in which $|\beps| \sim 10^3$ (see \emph{e.g.} \cite{Wu} and 
references therein).

On the contrary, the parameter $\epsilon'$ is entirely due to direct $CP$ 
violation, as can be seen explicitly by rewriting it as
\begin{equation} 
  \epsilon' = 
  \frac{i}{\sqrt{2}} \omega (1-\beps^2) 
  \frac{\xi_2-\xi_0}{(1+i\beps \, \xi_0)(1+i\beps \, \xi_2)}
\end{equation} 
which make evident that a (phase) difference of the weak decay amplitudes to 
two isospin channels is required to have $\epsilon' \neq 0$ (we remark in 
passing that a difference in the strong phases $\delta_I$ is not required to 
have $\epsilon' \neq 0$, since this parameter also gets a contribution from 
the interference of decays with and without mixing \cite{Cimento}).

In terms of the above parameters, the amplitude ratios for $\pi^+ \pi^-$ and 
$\pi^0 \pi^0$ decays are written respectively as
\begin{eqnarray}
  \eta_{+-} = \epsilon + \frac{\epsilon'}{1+\omega/\sqrt{2}} \quad && \quad 
  \eta_{00} = \epsilon - \frac{2\epsilon'}{1-\omega\sqrt{2}} 
\label{eq:eta}
\end{eqnarray}

\section{Alternative formulations}
\label{sec:others}

Several other definitions of the $CP$-violating parameters for the neutral 
$K$ system are used in past and recent literature: most of those coincide 
when suitable approximations are done; some of these approximations are 
physically justified in terms of small parameters, while others just depend on 
arbitrary phase convention choices.

Ignoring definitions which differ from the one described above just for 
trivial factors of $\sqrt{2}$, the most frequent cases found in the literature 
are the following \footnote{In listing the alternative definitions we always 
use the same symbols $\tilde{\epsilon}, \tilde{\omega}, \tilde{\epsilon'}$ in 
order to distinguish them from our standard definitions in eqs. (\ref{eq:eps}, 
\ref{eq:omega}, \ref{eq:epsprime}), but they clearly refer to different 
quantities in each case.}:
\begin{itemize}

\item
A minor variation \cite{Wolfenstein} is that in which only the definition of 
$\epsilon'$ is changed to
\begin{equation}
  \tilde{\epsilon}' \doteq 
  \frac{\omega}{\sqrt{2}} \left( i \, \xi_2 + \beps \right)
\end{equation}
so that 
\begin{eqnarray}
  \eta_{+-} = \frac{\epsilon + \tilde{\epsilon'}}{1+\omega/\sqrt{2}} &&
  \eta_{00} = \frac{\epsilon -2 \, \tilde{\epsilon}'}{1-\omega \sqrt{2}}
\label{eq:etawolf}
\end{eqnarray}

\item
Another definition \cite{Belusevic} is
\begin{equation}
  \tilde{\epsilon}' \doteq \frac{1}{\sqrt{2}} 
  \frac{\langle (\pi\pi)_{I=2} | \mathcal{T} | \kl \rangle}
       {\langle (\pi\pi)_{I=0} | \mathcal{T} | \ks \rangle} -
  \frac{i}{\sqrt{2}} \, \xi_0 \, \omega
\end{equation}
to which the definition in eq. (\ref{eq:epsprime}) reduces in the limit 
$|\beps| \ll 1$.

\item
Some authors \cite{Grimus} define the parameters in terms of 
the weak phases of $\kz, \kzb$ decay amplitudes into isospin eigenstates,  
\emph{i.e.}
\begin{eqnarray}
&& \tilde{\epsilon} \doteq \bar{\epsilon} + i \, \xi_0 \label{eq:epsgrimus} \\
&& \tilde{\omega} \doteq 
   \frac{\Rea(a_2)}{\Rea(a_0)} e^{i(\delta_2 - \delta_0)} \\
&& \tilde{\epsilon}' \doteq 
   \frac{i}{\sqrt{2}}\omega (\xi_2 - \xi_0)
\end{eqnarray}
The expressions for $\eta_{+-}$ and $\eta_{00}$ read in this case 
\begin{eqnarray}
  & \eta_{+-} = \tilde{\epsilon} + 
    \frac{\tilde{\epsilon}' - i\beps \, \tilde{\epsilon} \, 
    (\xi_0 + \xi_2 \, \tilde{\omega}/\sqrt{2})}
    {1+\tilde{\omega}/\sqrt{2} + 
    i\beps \, (\xi_0 +\xi_2 \, \tilde{\omega}/\sqrt{2})} \\
  & \eta_{00} = \tilde{\epsilon} -
    \frac{2\tilde{\epsilon}' - i\beps \, \tilde{\epsilon} \, 
    (\xi_0 - \xi_2 \, \tilde{\omega} \sqrt{2})}
    {1-\tilde{\omega}/\sqrt{2} +
    i\beps \, (\xi_0 - \xi_2 \, \tilde{\omega} \, \sqrt{2})}
\end{eqnarray}
and reduce to the ones in (\ref{eq:eta}) when terms of order 
$\omega \, \xi_I$ and $\omega^2 \, \epseps$ are neglected.

\item
Another widespread definition \cite{Sanda} \cite{Bigi} \cite{Isidori} is that
in terms of the ratios of amplitudes for $\ks,\kl$ decays into physical states 
\begin{eqnarray}
  && \tilde{\epsilon} \doteq (2\eta_{+-} + \eta_{00})/3 \label{eq:bigi1} \\
  && \tilde{\epsilon}' \doteq (\eta_{+-} - \eta_{00})/3 \label{eq:bigi2}
\end{eqnarray}
which is obviously tailored to get exactly 
\begin{eqnarray}
  \eta_{+-} = \tilde{\epsilon} + \tilde{\epsilon}' \quad && \quad 
  \eta_{00} = \tilde{\epsilon} - 2 \, \tilde{\epsilon}'  
\label{eq:expt}
\end{eqnarray}
The definitions based on eqs. (\ref{eq:expt}) are sometimes described as the 
``experimental'' ones for the $\epsilon$ and $\epsilon'$ parameters.

\item
Other definitions \cite{Barmin} are based on the decay amplitudes of the $CP$ 
eigenstates $K_1$ ($CP=+1$) and $K_2$ \mbox{($CP=-1$)}:
\begin{equation}
  \tilde{\epsilon}' \doteq \frac{1}{\sqrt{2}} 
  \frac{\langle (\pi\pi)_{I=2} | \mathcal{T} | K_2\rangle}
       {\langle (\pi\pi)_{I=0} | \mathcal{T} | K_1\rangle} = 
  \frac{\omega}{\sqrt{2}} \frac{\eta_2 - \beps}{1-\beps \, \eta_0}
\label{eq:barmin}
\end{equation}
This definition makes very explicit the meaning of $\tilde{\epsilon}'$ as 
direct $CP$ violation parameter, but its relations with the observable 
quantities become more complicated.
$\tilde{\epsilon}'$ in eq. (\ref{eq:barmin}) reduces to eq. 
(\ref{eq:epsprime}) in the limit in which $CP$ violation is small and 
$|\beps| \ll 1$; in this case 
the phase of $\tilde{\epsilon}'$ (assuming $CPT$) is exactly $\delta_2 - 
\delta_0 + \pi/2$, and eqs. (\ref{eq:etawolf}) are valid.

\item
The previous scheme can be generalized \cite{Palmer} introducing for each 
final state $f$ the quantities 
\begin{equation}
  \tilde{\epsilon}'_f \doteq 
    \frac{1-\overline{A}_f/A_f}{1+\overline{A}_f/A_f} = 
    \frac{\beps - \eta_f}{1-\beps \, \eta_f}
\label{eq:epspalmer}
\end{equation}
in analogy to the expression 
\begin{equation}
  \beps = \frac{1-q/p}{1+q/p}
\end{equation}
where 
\begin{eqnarray}
  p \doteq \langle \kz | \ks \rangle = (1+\beps) \quad && \quad 
  q \doteq \langle \kzb | \ks \rangle = (1-\beps) 
\end{eqnarray}

In any phase convention in which $|\beps| \ll 1$, the expression for $\eta_f$ 
reduces in this case to 
\begin{equation}
  \eta_f \simeq \beps + \tilde{\epsilon}'_f
\end{equation}
and for the $\pi\pi$ states, in the limit $|\omega| \ll 1$ 
\begin{eqnarray}
  \tilde{\epsilon}'_{+-} \simeq i \, \xi_0 + \epsilon' &&
  \tilde{\epsilon}'_{00} \simeq i \, \xi_0 - 2 \, \epsilon' 
\end{eqnarray}
giving back eqs. (\ref{eq:expt}).

\end{itemize}

We remind the reader that while both $\beps$ and the quantity in eq. 
(\ref{eq:epspalmer}) are not rephasing-invariant, the quantity
\begin{equation}
    \lambda_f \doteq \frac{1-q \overline{A}_f/p A_f}{1+q \overline{A}_f/p A_f} 
\end{equation}
commonly used in the phenomenological description of $CP$ violation in the 
$B$ system, has this property.
In terms of such parameter the quantities defined in eqs. (\ref{eq:eps}, 
\ref{eq:epsprime}) are 
\begin{eqnarray}
  \epsilon = \frac{1-\lambda_0}{1+\lambda_0} \quad && \quad 
  \epsilon' = \sqrt{2} \, e^{i(\delta_2-\delta_0)} \frac{a_2}{a_0}
    \frac{\lambda_0-\lambda_2}{(1+\lambda_0)^2}
\end{eqnarray}
while the expressions for the quantities (\ref{eq:bigi1}, \ref{eq:bigi2})
are more complicated and not very illuminating.

The actual direct $CP$ violation parameter which theorists have been trying to 
compute for a long time with different approaches (see \emph{e.g.} 
\cite{Theory} for a recent review on the theoretical status of $\epsilon'$ 
computations) is expressed by \cite{Buras}:
\begin{equation}
  \epsilon'_{TH} \doteq 
  \frac{i}{\sqrt{2}} e^{i(\delta_2-\delta_0)} 
  \frac{\Imm(a_0)}{\Rea(a_0)} \left( 
  \frac{\Imm(a_2)}{\Imm(a_0)} - \frac{\Rea(a_2)}{\Rea(a_0)} \right) 
\label{eq:theo}
\end{equation}

Indeed, as $CP$ violation is small, the definition in eq. (\ref{eq:epsprime}) 
reduces to eq. (\ref{eq:theo}) in any phase convention in which 
$|\beps| \ll 1$ neglecting terms of order $|\beps \, \xi_I|$, when 
\begin{eqnarray}
  & \omega \simeq e^{i(\delta_2-\delta_0)} \frac{\Rea(a_2)}{\Rea(a_0)} \\
  & \eta_I \simeq \beps + i \, \xi_I \\
  & \epsilon' \simeq \frac{i}{\sqrt{2}} e^{i(\delta_2-\delta_0)} 
  \frac{\Rea(a_2)}{\Rea(a_0)} ( \xi_2 - \xi_0 )
  \label{eq:epsapprox}
\end{eqnarray}
without any approximation based on the size of $|\omega|$ (in eq. 
(\ref{eq:epsapprox}) $|\xi_I| \ll 1$ was also assumed).

When neglecting $|\omega|$ and adopting the phase convention in which the 
dominant amplitude $a_0$ is real (the so-called Wu-Yang phase convention), one 
recovers the original expression of ref. \cite{WuYang}.

The PDG review on $CP$ violation \cite{PDG_CP} adopts the definitions of 
eqs. (\ref{eq:expt}) with eq. (\ref{eq:epsgrimus}), noting that one obtains 
eq. (\ref{eq:theo}) when terms of order $\epsilon' \, \Rea(a_2/a_0)$ are 
neglected.

\section{Consistency requirements}
\label{sec:consistency}

We would like to point out that not all the above definitions are 
consistent. In particular the so-called ``experimental'' expressions in 
eqs. (\ref{eq:expt}) can only be considered as the approximations of the exact 
eqs. (\ref{eq:eta}) for $|\omega| \ll 1$, and cannot be promoted to 
alternative \emph{definitions} of the $\epsilon$ and $\epsilon'$ parameters.

The reason is that there is an additional constraint that the amplitudes 
should satisfy, dictated by $CPT$ symmetry (which we have assumed throughout).
Ignoring electromagnetic effects (consistent with our neglecting of 
isospin-breaking effects), the $\pi\pi$ final states are not connected by 
strong interactions to other states: the $3\pi$ states for zero total angular 
momentum have opposite parity (conserved by strong interactions), and the 
$\pi\pi \gamma$ states require electromagnetism.
It follows that $CPT$ symmetry by itself requires the equality of partial 
decay rates for particle and antiparticle:
\begin{equation}
  \Gamma(\kz \To \pi^+ \pi^-) + \Gamma(\kz \To \pi^0 \pi^0) = 
  \Gamma(\kzb \To \pi^+ \pi^-) + \Gamma(\kzb \To \pi^0 \pi^0)
\end{equation}
This constraint can be expressed as a function of the physical decay 
amplitudes for $\ks\kl$ and the mixing parameters:
\begin{equation}
  |\langle \pi^+ \pi^- | \mathcal{T} | \ks \rangle|^2 \left[ 
  2\Rea(\eta_{+-}) - \langle \ks | \kl \rangle (1+|\eta_{+-}|^2) \right] +
  |\langle \pi^0 \pi^0 | \mathcal{T} | \ks \rangle|^2 \left[ 
  2\Rea(\eta_{00}) - \langle \ks | \kl \rangle (1+|\eta_{00}|^2) \right] = 0
\end{equation}
which is explicitly invariant for rephasing of the $|\kz\rangle, 
|\kzb\rangle$ states, since
\begin{equation}
  \langle \ks | \kl \rangle = \frac{2\Rea(\beps)}{1+|\beps|^2} = 
  \frac{2\Rea(\epsilon)}{1+|\epsilon|^2}
\end{equation}
Writing, without any loss of generality
\begin{eqnarray}
  \eta_{+-} = \epsilon + \epsilon_{+-} && 
  \eta_{00} = \epsilon + \epsilon_{00}
\end{eqnarray}
and using the isospin decomposition of the decay amplitudes (neglecting 
$|\Delta I| >3/2$ amplitudes and isospin-breaking effects), the constraint 
equation reduces to
\begin{eqnarray}
  |1+\omega/\sqrt{2}|^2 \left[ 
  2 \, \Rea(\epsilon_{+-}) - \langle \ks | \kl \rangle |\epsilon_{+-}|^2 -
  2 \, |\epsilon|^2 \langle \ks | \kl \rangle \, \Rea(\epsilon_{+-}/\epsilon) 
  \right] + \nonumber \\
  |1/\sqrt{2}-\omega|^2 \left[ 
  2 \, \Rea(\epsilon_{00}) - \langle \ks | \kl \rangle |\epsilon_{00}|^2 -
  2 \, |\epsilon|^2 \langle \ks | \kl \rangle \, \Rea(\epsilon_{00}/\epsilon) 
  \right] = 0
\end{eqnarray}
Since we know experimentally that $|\epsilon| = O(10^{-3})$ and 
$|\epsilon_{+-}|,|\epsilon_{00}| = O(10^{-6})$, we keep terms up to first 
order in $|\epsilon_{+-}|$ and $|\epsilon_{00}|$, obtaining
\begin{equation}
  |1+\omega/\sqrt{2}|^2 \, \Rea(\epsilon_{+-}) +
  |1/\sqrt{2}-\omega|^2 \, \Rea(\epsilon_{00}) = 0
\label{eq:midconstraint}
\end{equation}
Now, keeping only terms which are first order in $|\omega|$ and using the 
experimental fact that $\delta_2-\delta_0 \simeq -\pi/4$, one is finally
led to
\begin{equation}
  2 \, (1+|\omega|) \, \Rea(\epsilon_{+-}) + 
  (1-2|\omega|) \, \Rea(\epsilon_{00}) = 0
\label{eq:finalconstraint}
\end{equation}
It seems that such a constraint was not discussed in this context in the 
literature.

Clearly, this equation is trivially satisfied in absence of direct $CP$ 
violation, when $\epsilon_{+-} = \epsilon_{00} = 0$. 

In the stronger approximation in which all terms containing $|\omega|$ are 
neglected, the constraint becomes
\begin{equation}
  2 \, \Rea(\epsilon_{+-}) = -\Rea(\epsilon_{00})
\end{equation}
which is satisfied by the choice 
\begin{eqnarray}
  \epsilon_{+-} = \epsilon' \quad & \quad
  \epsilon_{00} = -2 \, \epsilon'
\end{eqnarray}
so that one gets back the approximate eqs. (\ref{eq:expt}).

Using instead the relations in eqs. (\ref{eq:eta}), consistent with the 
definitions in eqs. (\ref{eq:eps},\ref{eq:omega},\ref{eq:epsprime}), one 
has
\begin{eqnarray}
  \epsilon_{+-} = \frac{\epsilon'}{1+\omega/\sqrt{2}} \quad & \quad 
  \displaystyle
  \epsilon_{00} = \frac{-2 \, \epsilon'}{1-\omega \sqrt{2}}
\end{eqnarray}
for which the constraint of eq. (\ref{eq:midconstraint}) is equivalent to
\begin{equation}
  \Rea(\epsilon' \omega^*) = 0
\label{eq:epsconstraint}
\end{equation}
which is indeed satisfied at the level of approximation considered here:
writing the phase of $\epsilon'$ (defined in eq. (\ref{eq:epsprime})) in 
terms of the phase of $\omega$ as 
\begin{equation}
  \phi(\epsilon') = \phi(\omega) + \pi/2 + \delta \phi
\end{equation}
the constraint (\ref{eq:epsconstraint}) requires $\delta \phi = 0$ (mod $\pi$).
Since the exact expression for $\epsilon'$ is
\begin{equation}
  \epsilon' = \frac{i\omega}{\sqrt{2}} \left[(\xi_2 - \xi_0)
  \frac{1-\beps^2}{(1+i\, \beps \, \xi_0)(1+i\, \beps \, \xi_2)} \right]
\label{eq:epsprimeexact}
\end{equation}
$\delta \phi$ is the phase of the term in square brackets in the above 
expression (\ref{eq:epsprimeexact}), which can be seen to be indeed small by 
using the phase convention $|\beps| \ll 1$, since it is an invariant quantity 
under rephasing as can be easily verified. Its value is 
$\delta \phi \simeq 0.6 \cdot 10^{-5}$.

It should be mentioned that a phase space correction factor is required to 
account for the difference in the $\pi^\pm$ and $\pi^0$ masses when expressing 
the partial decay rates in terms of the amplitudes:
\begin{equation}
  a_{PS} = \frac{\sqrt{m(\kz)^2-4m(\pi^\pm)^2}}
                {\sqrt{m(\kz)^2-4m(\pi^0)^2}} \simeq 0.9855
\end{equation}
Strictly speaking, this 1.5\% effect should be neglected consistently 
in the exact isospin limit; partially accounting for isospin-breaking in this 
way, the constraint equation (\ref{eq:midconstraint}) is modified into 
\begin{equation}
  a_{PS} |1+\omega/\sqrt{2}|^2 \, \Rea(\epsilon_{+-}) +
  |1/\sqrt{2}-\omega|^2 \, \Rea(\epsilon_{00}) = 0
\label{eq:ibmidconstraint}
\end{equation}
The expressions in eqs. (\ref{eq:eta}) still satisfy this constraint at the 
same level of approximation as before: in this case instead of eq. 
(\ref{eq:epsconstraint}) one gets 
\begin{equation}
  (a_{PS}-1) \, \Rea(\epsilon') + 
  \frac{2+a_{PS}}{\sqrt{2}} \, \Rea(\epsilon' \omega^*) = 0
\label{eq:newconstraint}
\end{equation}
but the left-hand side of eq. (\ref{eq:newconstraint}) can be seen to be still 
proportional to $\delta \phi \simeq 0$.

Summarizing, the expressions in eqs. (\ref{eq:expt}) are approximations which 
are valid in the limit in which the parameter $|\omega|$ parameterizing the 
violation of the $\Delta I = 1/2$ rule is neglected, and cannot be considered 
as consistent alternative definitions of parameters describing $CP$ violation
in the kaon system.
It should be noted, furthermore, that the approximation in which $|\omega|$ is 
neglected is - strictly speaking - not a consistent one in this context, 
since if $\omega = 0$ the absence of $\Delta I = 3/2$ amplitudes would imply 
that no direct $CP$ violation is possible for neutral kaons decays into 
$\pi\pi$,due to lack of an amplitude interfering with the dominant 
($\Delta I = 1/2$) one.

\section{Experimental results}
\label{sec:averages}

In the experiments performed so far with neutral $K$ mesons, the information 
on direct $CP$ violation is extracted from the experimental measurement of the 
so-called ``double ratio'' $R$ of partial decay widths:
\begin{equation}
  R \doteq 
  \frac{\Gamma(\kl \rightarrow \pi^0 \pi^0)}
       {\Gamma(\ks \rightarrow \pi^0 \pi^0)}
  \frac{\Gamma(\ks \rightarrow \pi^+ \pi^-)}
       {\Gamma(\kl \rightarrow \pi^+ \pi^-)} = 
       \left| \frac{\eta_{00}}{\eta_{+-}} \right|^2
\end{equation}
This quantity is related to $\epsilon'/\epsilon$ by the following 
approximate expression 
\begin{equation}
  R \simeq 1 -  6 \Rea(\epsilon'/\epsilon) 
  -3\sqrt{2} \, \Rea \left( \omega^* \, \epsilon'/\epsilon \right)
\label{eq:rbest}
\end{equation}
in which second order terms in $\epsilon'$ or $\omega$ were neglected.
It is well known that the $\epsilon'/\epsilon$ ratio is close to being real,
since \cite{PDG2002} \cite{KTeV_eprime} (see also \cite{KLOE_delta})
\begin{eqnarray}
  && \phi(\epsilon) \simeq 2\Delta m /\Delta \Gamma = 
     (43.46 \pm 0.05)^\circ \\
  && \phi(\epsilon') \simeq \delta_2 - \delta_0 + \pi/2 = 
     (48 \pm 4)^\circ
\end{eqnarray}
(here, as usual, $\Delta m \doteq m(\kl)-m(\ks) >0$, $\Delta \Gamma \doteq 
\Gamma_S - \Gamma_L >0$) where the first approximate equality becomes exact 
in the limit in which the $\pi\pi$ decay amplitude dominates (always assuming 
$CPT$ symmetry), while the second one only depends on the smallness of $CP$ 
violation.
Equation (\ref{eq:rbest}) therefore reduces to \cite{Buras_omega}
\begin{equation}
  R \simeq 1 -  6 \, \Rea(\epsilon'/\epsilon) 
    \left[ 1 + \Rea(\omega)/\sqrt{2} \right]
\label{eq:rgood}
\end{equation}
which is commonly approximated to 
\begin{equation}
  R \approx 1 -  6 \, \Rea(\epsilon'/\epsilon) 
\label{eq:rbad}
\end{equation}
by neglecting $|\omega|$.
Equation (\ref{eq:rbad}) is the one routinely used in the experimental papers.

Since $|\omega|$ is of order 5\%, the difference between using eq. 
(\ref{eq:rgood}) and (\ref{eq:rbad}) amounts to a reduction of the value of 
$\Rea(\epsilon'/\epsilon)$ by 2.2\%, which is small when compared to the 
current precision of the theoretical computations, and also to the present 
experimental error (but not to the size of the systematic corrections 
applied by the experiments to obtain the central value). 
Extrapolating to a final experimental precision of $1 \cdot 10^{-4}$ on 
$\Rea(\epsilon'/\epsilon)$ in a few years from now, the use of the correct 
expression eq. (\ref{eq:rgood}) will be appropriate.

Averaging the most precise results on $\Rea(\epsilon'/\epsilon)$ at face value 
one obtains $\Rea(\epseps) = (16.7 \pm 2.3) \cdot 10^{-4}$ where the error has 
been inflated by a factor 1.44 according to the procedure adopted by the PDG 
\cite{PDG2002}, due to the poor $\chi^2$ value of 6.2 (with 3 degrees of 
freedom). 

\begin{figure}[hbt!]
\begin{center}
  \includegraphics[scale=0.5]{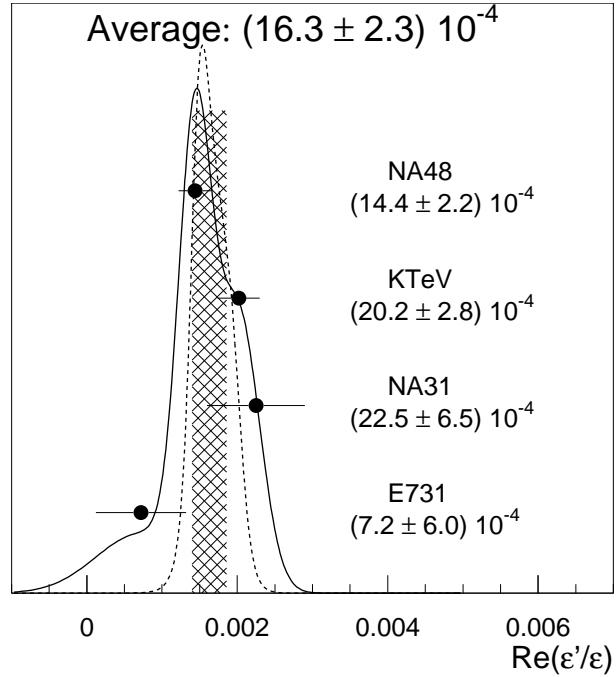}
    \caption{Ideogram of recent published $\Rea(\epseps)$ measurements as 
corrected according to eq. (\ref{eq:rgood}).
The curves show (unnormalized) probability distributions according to the PDG 
procedure \cite{PDG2002} (solid line) or a Bayesian ``skeptical'' approach 
\cite{D'Agostini} (dashed line).}
    \label{fig:ideo}
\end{center}
\end{figure}

Using eq. (\ref{eq:rgood}) one obtains instead: 
\begin{equation}
  \Rea(\epseps) = (16.3 \pm 2.3) \cdot 10^{-4}
\end{equation}
as the value to be compared to theoretical computations, and the $\chi^2$ 
improves only marginally to 5.9, without affecting the scaled error in a 
significant way. 
A graphical depiction of the present data is shown in figure \ref{fig:ideo}. 
The probability of the four most precise measurements to be consistent is 
11\%, varying between 9\% and 22\% when a single measurement is ignored. 

It should also be noted that the uncertainty on the value of $|\omega|$ 
hardly affects any comparison with theory in itself, since in any case the 
empirical value of such parameter is used both in the computation (due to 
the theoretical difficulties with the $\Delta I = 1/2$ rule) and in 
extracting the value of $\Rea(\epseps)$ from the experiments.

The similarity of the phase of $\epsilon'$ with that of $\epsilon$ is 
an accidental fact which hinges on the validity of the $CPT$ symmetry, without 
which the phase of $\epsilon$ would be different from the ``super-weak'' value
$2 \, \Delta m/\Delta \Gamma$; for this reason the smallness of 
$\Imm(\epsilon'/\epsilon)$ is considered a test of such symmetry.
The difference of such phases is however experimentally constrained 
\cite{KTeV_eprime} to be tiny: $\phi(\epsilon') - \phi(\epsilon) = 
(-1.2 \pm 1.5)^\circ$, and therefore the use of eq. (\ref{eq:rbest}) is not 
required; experiments usually assume $CPT$ symmetry explicitly 
\cite{KTeV_eprime} or implicitly \cite{NA48_eprime} in the extraction of 
$\epsilon'/\epsilon$.

It should be reminded that $\Imm(\epsilon'/\epsilon)$ can be measured using 
kaon interferometry \cite{Imeps} \cite{Sanda}, and would be therefore 
accessible to the KLOE experiment \cite{KLOE_Im} when a sufficient statistics 
will be accumulated.

\section{Conclusions}
\label{sec:conclusions}

In view of the recent and future progress, in both experiment and theory, in 
the determination of the parameter $\epsilon'$ measuring direct $CP$ violation 
in neutral kaon decays, the use of a common definition for it is advisable.
We reviewed some of the choices present in the literature, showing that the 
simple so-called ``experimental'' one is necessarily an approximation, which 
is still good at the present level of accuracy but would have to be abandoned 
in the future to allow an accurate comparison of theory and experiment.

\begin{acknowledgments}
This work originated from one of the several seemingly casual - but always 
profound - remarks by I. Mannelli, which the author is pleased to thank.
We also thank G. Isidori for his comments on the draft of this paper.
\end{acknowledgments}

\bibliography{epspaper}

\end{document}